\documentclass[12pt]{article}

\begin{document}

\begin{center}

{\Large {Simple way to calculate UV-finite one-loop quantum energy in Randall-Sundrum model}}

\vspace{1,5cm}

{Boris L. Altshuler}\footnote{E-mail addresses: baltshuler@yandex.ru $\,\,\,  \& \,\,\,$  altshul@lpi.ru}

\vspace{1cm}

{\it Theoretical Physics Department, P.N. Lebedev Physical
Institute, \\  53 Leninsky Prospect, Moscow, 119991, Russia}

\vspace{1,5cm}

\end{center}

{\bf Abstract:} The surprising simplicity of Barvinsky-Nesterov or equivalently Gelfand-Yaglom methods of calculation of quantum determinants permits to obtain compact expressions for UV-finite difference of one-loop quantum energies for two arbitrary values of parameter of the double-trace asymptotic boundary conditions. This result generalizes Gubser and Mitra calculation for particular case of difference of "regular" and "irregular" one-loop energies in one-brane RS-model. Approach developed in the paper also allows to get "in one line" the one-loop quantum energies in two-brane RS-model. The relationship between "one-loop" expressions corresponding to mixed Robin and to double-trace asymptotic boundary conditions is traced.

\vspace{0,5cm}


\newpage

\section{Introduction}

\quad In 2001 Witten \cite{Witten} showed that in frames of the AdS/CFT correspondence multi-trace deformation $W(\hat{O})$ of the boundary quantum field theory may be equivalent to the boundary condition ({\it{further on}} - b.c.)

\begin{equation}
\label{1}
\alpha = \frac{\partial W(\beta)} {\partial \beta}
\end{equation}
imposed upon the regular ($\alpha$) and irregular ($\beta$) asymptotics at the AdS horizon ($z \to 0$) of the bulk scalar field $\phi$:

\begin{equation}
\label{2}
\phi = \alpha\,z^{\frac{d}{2}+\nu} + \beta\,z^{\frac{d}{2} - \nu},
\end{equation}
where $\alpha$ corresponds to the source of single-trace operator $\hat{O}$ whereas $\beta$ - to its quantum average.

In case of the double-trace deformation $W = (1/2)f \hat{O}^{2}$ (\ref{1}) comes to

\begin{equation}
\label{3}
\alpha = f\,\beta.
\end{equation}

Here Euclidean metric of $(d+1)$ dimensional AdS space of the Randall-Sundrum (RS) model is taken in a form

\begin{equation}
\label{4}
ds^{2}= \frac{dz^{2}+\eta_{\mu\nu}dx^{\mu}dx^{\nu}}{(kz)^{2}},
\end{equation}
and $\epsilon < z < L$ ($z=\epsilon, \, L$ are position of UV and IR branes), $\mu, \nu = 0,1,...(d-1)$, $\eta_{\mu\nu}=\delta_{\mu\nu}$ in Euclidean signature, $k$ is AdS curvature scale, and $\phi=\phi(\vec{p},z)$ satisfy the equation ($\vec{p}$ is momentum in Euclidean $d$-space, $p=|\vec{p}|$):

\begin{equation}
\label{5}
\hat{D}(p) \phi = \left[-z^{2}\frac{\partial^{2}}{\partial z^{2}}+(d-1)z\frac{\partial}{\partial z}+\left(\nu^{2}-\frac{d^{2}}{4}\right)+z^{2} p^{2}\right]\,\phi = 0,
\end{equation}
\\
$\nu = \sqrt{d^{2}/4 + m^{2}/k^{2}}$ for minimal action of the bulk scalar field of mass $m$.

Gubser and Mitra showed in \cite{Mitra} (see also \cite{Hartman}, \cite{Diaz}) that difference of bulk Green functions satisfying asymptotic b.c. (\ref{3}) for two values of double-trace parameter $f$ is UV-finite at coinciding arguments:

\begin{equation}
\label{6}
\int [G_{f_{2}}(p;z,z) - G_{f_{1}}(p;z,z)]\,d^{d}p  \, < \, \infty,
\end{equation}
where Green function $G_{f}(p;z,z')$ is taken in Euclidean signature and is given by formula (32) of \cite{Mitra}:

\begin{eqnarray}
\label{7}
G_{f}(p;z,z')= \qquad \qquad  \qquad \qquad  \qquad \nonumber \\
- \frac{k^{d-1}(zz')^{d/2}K_{\nu}(pz')}{1+\bar{f}}\,
\{[I_{-\nu}(pz)+\bar{f}I_{\nu}(pz)]\,\theta(z'-z)+ (z \leftrightarrow z')\} \nonumber
\\
\\
\bar{f} = f\,\left(\frac{2}{p}\right)^{2\nu}\,\frac{\Gamma (1+\nu)}{\Gamma (1-\nu)}, \qquad  \qquad  \qquad  \qquad \qquad \nonumber
\end{eqnarray}
$I_{\pm \nu}$, $K_{\nu}$ are Bessel functions of imaginary argument, here $L=\infty$, and expression for $\bar{f}$ is obtained from comparison of asymptotic of $(I_{-\nu}+\bar{f}I_{\nu})$ at $z \to 0 $ with (\ref{2}), (\ref{3}) \cite{Mitra}.

However, as it was pointed out in \cite{Mitra}, it is hard to calculate for general values of $f$ the one-loop vacuum energy corresponding to difference (\ref{6}) of Green function.

In the present paper, which is development of \cite{Alt2015}, calculation of this one-loop energy is performed with simple "boundary operator" formula, proposed by Barvinsky and Nesterov (B-N) \cite{Barv2005}-\cite{Barv2014} for ratio of determinants of one and the same differential operator in one-dimensional problem for two different b.c. imposed upon eigenfunctions of $\hat{D}$. It will be shown also that in this case B-N approach is equivalent to Gelfand-Yaglom (G-Y) method \cite{GY}-\cite{Kirsten}.

Let us describe in  short B-N and Y-M approaches which application is crucial for this paper.

As it was demonstrated in \cite{Barv2005}-\cite{Barv2014} the ratio of determinants of differential operator for two different b.c. is equal to ratio of determinants of certain "boundary operators" given by the corresponding Green functions with their arguments taken at the boundary. The idea behind it is seemingly simple although proves to be very effective: Gauss functional integral, which gives the looked for determinant, is a product of the functional integral over bulk field with fixed values at the boundaries (that is when Dirichlet b.c. are imposed) and of the functional integral over boundary values of the field weighted by boundary operator depending on b.c. under consideration; thus in the ratio of determinants bulk functional integrals reduce. In one-dimensional problem boundary is a dot and boundary operator is just a number equal, as it is shown in \cite{Barv2005}-\cite{Barv2014}, to the value of corresponding Green function at the boundary. Finally ratio of determinants comes to the product of ratios of boundary operators of the one-dimensional problem over quantity parameterizing one-dimensional problem (momentum $\vec{p}$ in transverse $d$-space in this paper).

G-Y approach \cite{GY}-\cite{Kirsten} says that product of eigenvalues (determinant) of differential operator of the one-dimensional problem $\hat{D}\phi_{n}(z) = \lambda_{n}\phi_{n}(z)$ defined on interval $a<z<b$ and determined by b.c. $A[\phi(a)] = 0$ and $B[\phi(b)]=0$ ($A[\phi]$, $B[\phi]$ are some combinations of $\phi$ and its derivative $\phi'$ taken at corresponding points) may be expressed through solution $v(z)$ of homogeneous equation $\hat{D}v(z) = 0$ which obey given b.c. at one boundary, say at $z=a$, that is $A[v(a)]=0$; then G-Y method gives ${\rm Det}\hat{D} \sim B[v(b)]$. The logic of the proof of this quite effective formula is double-step: (1) for solution $\phi(z|\lambda)$ of Eq. ${\rm Det}\hat{D}\phi = \lambda\,\phi$, which obey b.c. $A[\phi(a|\lambda)] = 0$ and which is considered as a function of $\lambda$, function $B(\lambda) \equiv B[\phi(b|\lambda)]$ have zeroes at $\lambda = \lambda_{n}$; (2) since logarithmic derivative of $B(\lambda)$ ($d\ln{B(\lambda)}/d\lambda$) has poles in complex $\lambda$-plane exactly at $\lambda = \lambda_{n}$ it is possible to express $\zeta$-function ($\zeta (s) = \sum{\lambda_{n}}^{-s}$) with contour integral over this logarithmic derivative and finally, after a number of rather conventional steps, to get the looked for G-Y formula $e^{-\zeta'(0)}={\rm{Det}}\hat{D} \sim B(\lambda=0) = B[\phi(b|0)] = B[v(b)]$ (since $\phi(z|0)$ is nothing but a homogeneous solution $v(z)$ introduced above in this paragraph).

As to our knowledge the correspondence of B-N and G-Y methods was not considered in literature so far. The bulk of the paper consists of the examples of application of B-N method with certain parallels with G-Y approach. In Appendix the power of G-Y method is demonstrated by a number of physical problems where G-Y formulas immediately give well known values of Casimir potential calculated conventionally in a rather complex way.

In standard approach applied in \cite{Mitra} - \cite{Alt2015} calculation of one-loop energy $V^{(d)}$ in $(d+1)$ dimensional RS-model is performed with three integrations: over $p$ like in (\ref{6}), over $z$ between its endpoints, and over mass squared parameter $\alpha$ according to the well known identity:

\begin{equation}
\label{8}
V= \frac{1}{2}\,{\rm ln}\,{\rm Det}\,\hat{D} = \int d\tilde{\alpha}\frac{\partial V}{\partial \tilde{\alpha}} = \frac{1}{2} \int^{\alpha}d\tilde{\alpha}\,{\rm Tr}\,{\rm ln}\,G(x,z;x,z;\tilde{\alpha}).
\end{equation}

B-N or G-Y methods permit to "jump over" integrations over $z$ and $\alpha$, and immediately give answer for ratio of determinants of differential operator of one-dimensional problem parametrized in our case by $p$ (see (\ref{5})). Then corresponding difference of one-loop quantum energies in $d$ dimensions is given by integral over $\vec p$:

\begin{equation}
\label{9}
V^{(d)}_{2}-V^{(d)}_{1} = \frac{1}{2}\int\,\frac{d^{d}p}{(2\pi)^{d}}\, \ln \left[\frac{{\rm Det}_{2}\hat{D}(p)}{{\rm Det}_{1}\hat{D}(p)}\right].
\end{equation}
It is shown in the paper that integral in (\ref{9}) is UV-convergent if indexes $2,1$ in (\ref{9}) refer to two values $f_{2},\, f_{1}$ of the double-trace parameter in asymptotic b.c. (\ref{3}). And on the other hand integral in (\ref{9}) is UV-divergent if these indexes refer to two fixed Robin parameters of mixed b.c. imposed at $z=\epsilon$.

The "strange discrepancy" (see Sec. III in \cite{Alt2015}) of expressions for difference of "regular" and "irregular" one-loop energies $(V_{+}-V_{-}=V_{f=\infty}-V_{f=0})$ calculated with different choice of parameter $\alpha$ in (\ref{8}) ($\sqrt{\alpha} = m$ in \cite{Mitra}-\cite{Diaz}, and $\sqrt{\alpha}$ is auxiliary mass introduced in \cite{Alt2015}) perhaps is resolved in this paper. In any case formula for $V^{(d)}_{+}-V^{(d)}_{-}$ obtained in section 3 differs from both competing expressions of \cite{Mitra} and \cite{Alt2015}.

Structure of the paper is as follows. In Sec. 2 the work of B-N and G-Y methods, their equivalence, and correspondence of Robin b.c. and asymptotic b.c. are demonstrated by an elementary dynamical example. Sec. 3 presents results of calculation of UV-finite one-loop quantum energy for double-trace asymptotic b.c. in one-brane ($L = \infty$) and two-branes ($L < \infty$) RS-models. Conclusion outlines the possible ways of future work. Appendix presents a number of striking examples of power of G-Y method.

\section{Elementary example: identity of B-N and G-Y methods}

\quad In this Section we demonstrate the identity of Barvinsky-Nesterov (B-N) and Gelfand-Yaglom (G-Y) methods of calculation of ratio of quantum determinants determined by different Robin or asymptotic b.c. on an elementary example of differential operator $\hat{D}_{0}$ of massless scalar field $\phi$ in flat $(d+1)$ dimensions:

\begin{equation}
\label{10}
\hat{D}_{0}(p) \,\phi (p,z) = \left[-\frac{\partial ^{2}}{\partial z^{2}} + p^{2}\right]\,\phi (p,z).
\end{equation}
$\hat{D}_{0}$ is defined on interval $\epsilon <z< L$ (here are kept the notations used in the Introduction).

We consider two spectra of eigenvalues $\lambda^{(1)}_{n}$, $\lambda^{(2)}_{n}$ of Eq. $\hat{D}_{0}\phi_{1,2}(z) = \lambda \phi_{1,2}(z)$ determined by one and the same Neumann b.c. at $z=L$ and two mixed Robin b.c. at $z=\epsilon$ (prime means derivative over $z$ throughout the paper):

\begin{equation}
\label{11}
\phi'_{1,2}(L)=0; \,\, \, \phi'_{1,2}(\epsilon) + r_{1,2}\phi_{1,2}(\epsilon) = 0.
\end{equation}
Then, according to B-N boundary operator approach, the ratio of corresponding determinants of $\hat{D}_{0}$ is given by the ratio of corresponding Green functions with their both arguments taken at the boundary where different b.c. are imposed (that is at $z=\epsilon$ in our example):

\begin{equation}
\label{12}
\frac{\prod_{n} \lambda_{n}^{(2)}}{\prod_{n} \lambda_{n}^{(1)}} = \frac{{\rm Det}_{r_{2}-N}\hat{D}_{0}(p)}{{\rm Det}_{r_{1}-N}\hat{D}_{0}(p)} = \left.\frac{G^{(0)}_{r_{1}-N}(p;z,z')}{G^{(0)}_{r_{2}-N}(p;z,z')}\right|_{z=z'=\epsilon} \equiv Q_{0}(p).
\end{equation}

Green functions in (\ref{12}) obeying Eq. $\hat{D}_{0}(p)G^{(0)}(p;z,z') = \delta (z-z')$ and b.c. (\ref{11}) are given by standard expression:

\begin{equation}
\label{13}
G^{(0)}_{r-N}(p;z,z') = \frac{u_{r}(z)\,v(z')\,\theta (z'-z) + (z \leftrightarrow z')}{u'_{r}v - u_{r}v'},
\end{equation}
where $v(z) = \cosh p(z-L)$ and $u_{r_{1,2}}(z) = p\cosh p(z-\epsilon) - r_{1,2} \sinh p(z-\epsilon)$ obey b.c. (\ref{11}):

\begin{equation}
\label{14}
v'(L)=0; \,\, \, u'_{r_{1}}(\epsilon) + r_{1} u_{r_{1}}(\epsilon) = 0; \,\,\, u'_{r_{2}}(\epsilon) + r_{2} u_{r_{2}}(\epsilon) = 0.
\end{equation}
Thus for the ratio of determinants $Q_{0}(p)$ (\ref{12}) it is obtained from (\ref{13}) and from the explicit expressions for $v(z)$ and $u_{r_{1,2}}(z)$:

\begin{equation}
\label{15}
Q_{0}(p) = \frac{u_{r_{1}}(\epsilon)}{u_{r_{2}}(\epsilon)} \cdot \frac{u'_{r_{2}}v - u_{r_{2}}v'}{u'_{r_{1}}v - u_{r_{1}}v'} = \frac{p\,\sinh p(L-\epsilon)-r_{2}\,\cosh p(L-\epsilon)}{p\,\sinh p(L-\epsilon)-r_{1}\,\cosh p(L-\epsilon)}.
\end{equation}

Let us demonstrate now the identity of B-N expressions (\ref{12}), (\ref{15}) with the Gelfand-Yaglom (G-Y) formula for the ratio of determinants:

\begin{equation}
\label{16}
Q_{0}(p) = \frac{{\rm Det}_{r_{2}-N}\hat{D}_{0}(p)}{{\rm Det}_{r_{1}-N}\hat{D}_{0}(p)} = \frac{v'(\epsilon) + r_{2} v(\epsilon)}{v'(\epsilon) + r_{1} v(\epsilon)},
\end{equation}
where $v(z) = \cosh p(z-L)$ is the introduced above solution of homogeneous equation $\hat{D}_{o}v(z)=0$ obeying Neumann b.c. at $z=L$.

The identity of expressions (\ref{15}) and (\ref{16}) is immediately seen from the explicit expression for $v(z)$ and also in general if we substitute in (\ref{15}) $u'_{r_{1,2}}(\epsilon) = - r_{1,2} u_{r_{1,2}}(\epsilon)$ from b.c. (\ref{14}). This simple observation proves to be quite useful in subsequent analysis.

Difference of one-loop energies corresponding to ratio of determinants (\ref{12})

\begin{equation}
\label{17}
V_{r_{2}}^{(d)} - V_{r_{1}}^{d} = \frac{1}{2}\,\int\frac{d^{d}p}{(2\pi)^{d}}\,\ln\,Q_{0}(p)
\end{equation}
is UV-divergent if Robin coefficients $r_{1}$, $r_{2}$ are fixed constants, as it is seen from explicit dependence $Q_{0}(p)$ given in (\ref{15}). We shall show however that application of this logic to asymptotic b.c. (\ref{3}) makes $r_{1}$, $r_{2}$ in (\ref{15}) dependent on $f_{1}$, $f_{2}$ and on momentum $p$ in a way that makes $Q_{0}(p) \to 1$ at $p \to \infty$, hence integral in (\ref{17}) is UV-finite in this case.

Analogy of asymptotic, at $z \to 0$, expression (\ref{2}) for elementary differential operator (\ref{10}) is $\phi = \alpha z + \beta$ (this formally corresponds to $d=1$, $\nu = 1/2$ in (\ref{2})). And analogy of Gubser-Mitra Euclidean Green function (\ref{7}) (although here we take $L < \infty$) obeying Neumann b.c. at $z=L$ and double-trace asymptotic b.c. $\alpha = f \beta$ at $z \to 0$ is Green function:

\begin{eqnarray}
\label{18}
G^{(0)}_{f-N}(p;z,z') = \frac{u_{f}(z)\,v(z')\,\theta (z'-z) + (z \leftrightarrow z')}{u'_{f}v - u_{f}v'} = \quad \quad \nonumber \\
\\
\frac{[\cosh pz + \bar{f} \sinh pz]\,\cosh p(z'-L)\,\theta(z'-z) + (z \leftrightarrow z')}{p\,(\sinh pL + \bar{f}\,\cosh pL)}; \nonumber \\
\nonumber  \\
\bar{f} = \frac{f}{p}. \nonumber \quad  \quad  \quad  \quad  \quad   \quad  \quad \quad \quad
\end{eqnarray}

Now, according to B-N prescription, we take the ratio of two Green functions (\ref{18}) for two double-trace parameters $f_{1}$, $f_{2}$ at $z=z'=\epsilon$ and, following (\ref{12}), $define$ with this ratio the ratio of corresponding determinants:

\begin{eqnarray}
\label{19}
\left.\frac{G^{(0)}_{f_{1}-N}(p;z,z')}{G^{(0)}_{f_{2}-N}(p;z,z')}\right |_{z=z'=\epsilon} = \frac{\cosh p\,\epsilon + \bar{f}_{1} \sinh p\,\epsilon}{\sinh p\,L + \bar{f}_{1} \cosh p\,L} \cdot \frac{\sinh p\,L + \bar{f}_{2}\,\cosh p\,L}{\cosh p\,\epsilon + \bar{f}_{2} \sinh p\,\epsilon} = \nonumber \\
\nonumber \\
\frac{{\rm Det}_{f_{2}-N}\hat{D}_{0}(p)}{{\rm Det}_{f_{1}-N}\hat{D}_{0}(p)} = \frac{\prod_{n} \tilde{\lambda}_{n}^{(2)}}{\prod_{n} \tilde{\lambda}_{n}^{(1)}} \equiv \tilde{Q}_{0} (p). \quad  \quad  \quad  \quad
\end{eqnarray}

Ratio (\ref{19}) depends on $\epsilon$ which is not present in definition of Green function (\ref{18}), as well as it is not present in (\ref{7}). As it was noted formula (\ref{19}) is actually a definition of the ratio of determinants, that is a definition of corresponding eigenvalues $\tilde{\lambda}_{n}^{(1), (2)}$ - just like authors of paper \cite{Mitra} $defined$ UV-finite one-loop energy $V_{+}^{(d)} - V_{-}^{(d)}$ with integral over $z$ from $\epsilon$ to $\infty$ although asymptotic b.c. (\ref{3}) is imposed at $z \to 0$ and integrand (which is the difference of regular and irregular Green functions) does not know anything about $z=\epsilon$.

Difference of vacuum energies corresponding to ratio of determinants (\ref{19}) and given by $\int d^{d}p\,\ln \tilde{Q}_{0}(p)$ (cf. (\ref{9}) or (\ref{17})) is UV-finite since for $\bar{f}$ weakly depending on $p$ (like in (\ref{18}) (c.f. also (\ref{7})) $\tilde{Q}_{0}(p)$ in (\ref{19}) $\to 1$ at $p \to \infty$.

There is the question: what Robin b.c. at $z=\epsilon$ characterized by parameter $r$ (like in (\ref{11})) corresponds to asymptotic condition (\ref{3}) characterized by double-trace parameter $f$? Or in other words: what are conditions of identity of $Q_{0}(p)$ and $\tilde{Q}_{0}(p)$ in the RHS of (\ref{12}) and (\ref{19}) correspondingly? Function $u_{f}(z) = \cosh pz + \bar{f}\,\sinh pz$ in expression for Green function (\ref{18}) formally obeys at $z=\epsilon$ Robin b.c. $u'(\epsilon) + r_{\epsilon}\,u(\epsilon) = 0$ for Robin parameter:

\begin{equation}
\label{20}
r_{\epsilon} = - \frac{u'(\epsilon)}{u(\epsilon)} = - \frac{p\,(\sinh p\,\epsilon + \bar{f}(p)\,\cosh p\,\epsilon)}{\cosh p\,\epsilon + \bar{f}(p)\,\sinh p\,\epsilon},
\end{equation}
and it is easy to check that substitution of (\ref{20}) in B-N ratio (\ref{15}) gives B-N ratio (\ref{19}) identically.

Also knowledge of $r_{\epsilon}=r_{\epsilon}(\bar{f})$ (\ref{20}) permits to put down the equations for spectra $\tilde{\lambda}_{n}^{(1),(2)}$ defined in (\ref{19}):

\begin{equation}
\label{21}
\sqrt{\tilde{\lambda}_{n}-p^{2}}\,\tan[\sqrt{\tilde{\lambda}_{n}-p^{2}}\,(L-\epsilon)] = -r_{\epsilon} = \frac{p\,(\tanh p\,\epsilon + \bar{f}(p))}{1 + \bar{f}(p)\,\tanh p\,\epsilon},
\end{equation}
$\bar{f}(p)$ see in (\ref{18}). This equation is obtained from spectral equation $\hat{D}\phi_{n} = \lambda_{n}\,\phi_{n}$, and b.c. (\ref{11}) where $r_{\epsilon}$ is taken from (\ref{20}). At $\epsilon = 0$ (\ref{21}) simplifies and also makes sense, as well as the ratio of determinants (\ref{19}) makes sense in the limit $\epsilon \to 0$. However in this case one-loop energy given by $\int d^{d}p\,\ln \tilde{Q}_{0}(p)$ is UV-divergent. Thus $\epsilon > 0$ really serves the UV-regulator of quantum loops in d-space; is not it curious to see this well know fact of AdS/CFT correspondence in the simplest example of this section.

Transcendental equation (\ref{21}) for $\tilde{\lambda}_{n}$ is valid in particular for 'regular' ($\bar{f} = \infty$) and 'irregular' ($\bar{f} = 0$) asymptotics. It is also seen that (\ref{21}) comes to spectral conditions for Neumann($\epsilon$)-Neumann($L$) or Dirichlet($\epsilon$)-Neumann($L$) b.c. for negative values of $\bar{f}$: $\bar{f}= - \tanh p\,\epsilon$ ($r_{\epsilon}=0$) and $\bar{f} = - 1/\tanh p\,\epsilon$ ($r_{\epsilon}=\infty$) correspondingly. Surely it is a sort of miracle that B-N approach gives simple expression (\ref{19}) for the ratio of infinite products of rather complex eigenvalues - solutions of equation (\ref{21}).

\section{One-loop quantum energy for asymptotic b.c. in RS-model: $L=\infty$ and $L < \infty$}

\subsection{One-brane RS-model}

In parallel with elementary example of section 2 we apply the B-N prescription \cite{Barv2005}-\cite{Barv2014} for calculation of the ratio of determinants of operator $\hat{D}(p)$ (\ref{5}) defined like in \cite{Mitra} for zero b.c. at IR infinity ($L=\infty$) and for two double trace asymptotics (\ref{3}). Like in (\ref{12}) the ratio of determinants is equal to the ratio of Green functions (\ref{7}) taken at $z=z'=\epsilon$:

\begin{equation}
\label{22}
\frac{{\rm Det}_{f_{2}}\hat{D}(p)}{{\rm Det}_{f_{1}}\hat{D}(p)} = \frac{G_{f_{1}}(p;\epsilon,\epsilon)}{G_{f_{2}}(p;\epsilon,\epsilon)} = \frac{I_{-\nu}(p\,\epsilon) + \bar{f}_{1}(p)I_{\nu}(p\,\epsilon)}{I_{-\nu}(p\,\epsilon) + \bar{f}_{2}(p)I_{\nu}(p\,\epsilon)}\cdot \frac{1 + \bar{f}_{2}(p)}{1 + \bar{f}_{1}(p)} \equiv Q(p,\epsilon).
\end{equation}
$\bar{f}_{1,2}$ are defined in (\ref{7}). For regular ($f_{2} = \infty$) and irregular ($f_{1} = 0$) asymptotics (\ref{3}) ratio of corresponding determinants (\ref{22}) is equal to $I_{-\nu}(p\epsilon)\,/\,I_{\nu}(p\epsilon)$. This was the result of "Remark B" in Conclusion of \cite{Alt2015}.

For $\bar{f}_{1(2)}$ given in (\ref{7}) $Q(p,\epsilon) \to 1$ at $p \to \infty$ (like $Q_{0}$ in (\ref{19})). Then one-loop energy corresponding to the ratio of determinants (\ref{22}) is UV-finite:

\begin{equation}
\label{23}
V^{(d)}_{f_{2}} - V^{(d)}_{f_{1}} = \frac{1}{2}\,\int\frac{d^{d}p}{(2\pi)^{d}}\,\ln\,Q < \infty.
\end{equation}
This conclusion is not valid for $\epsilon = 0$ in (\ref{22}) that is in absence of UV-brane screening AdS horizon. Thus here again - like in simple example of section 2 (cf. (\ref{19})) - $\epsilon$ plays a role of UV-regulator of UV divergencies of one-loop vacuum energy (\ref{23}).

In paper \cite{Mitra} $f_{2}=f$ and $f_{1}=0$ (irregular asymptotic b.c. denoted by index "-") were considered. And from (\ref{22}), (\ref{23}) it follows:

\begin{equation}
\label{24}
\tilde{V}^{(d)}(f) \equiv V^{(d)}_{f} - V^{(d)}_{-} = - \frac{\Omega_{d-1}}{2(2\pi)^{d}\epsilon^{d}}\,\int_{0}^{\infty}\,y^{d-1}dy\,\ln\,\left[\frac{I_{-\nu}(y) + \bar{f}(y,\epsilon)\,I_{\nu}(y)}{I_{-\nu}(y) (1+\bar{f}(y,\epsilon))}\right],
\end{equation}
where $y=p\,\epsilon$, $\Omega_{d-1}$ is volume of $(d-1)$-sphere of unit radius, and function $\bar{f}(y,\epsilon)$ in (\ref{24}) is easily seen from definition of $\bar{f}(p)$ in (\ref{7}): $\bar{f}(y,\epsilon) = f\,(2\epsilon)^{2\nu}\Gamma(1+\nu)\,/\,y^{2\nu} \Gamma(1-\nu)$. Thus potential (\ref{24}) is actually a function of dimensionless double-trace parameter $f\,\epsilon^{2\nu}$.

Formula for difference of regular and irregular one-loop energies $V^{(d)}_{+} - V^{(d)}_{-}$ follows from (\ref{24}) when $f=\infty$:

\begin{eqnarray}
\label{25}
V^{(d)}_{+} - V^{(d)}_{-} = \frac{\Omega_{d-1}}{2(2\pi)^{d}\epsilon^{d}}\,\int_{0}^{\infty}\,y^{d-1}dy\,\ln\left[\frac{I_{-\nu}(y)}{I_{\nu}(y)}\right] = \nonumber \\
\nonumber  \\
\frac{2\sin(\pi\nu)\Omega_{d-1}}{(2\pi)^{d+1}d\,\epsilon^{d}} \int_{0}^{\infty}\frac{y^{d-1}dy}{I_{\nu}(y)\,I_{-\nu}(y)}. \quad   \quad
\end{eqnarray}
This expression differs from the ones, also different, received for $V_{+}-V_{-}$ with standard procedure (\ref{8}) in \cite{Mitra}-\cite{Diaz} and in \cite{Alt2015}. The visible drawback of formulas (\ref{23}) - (\ref{25}) is in their zero value for integer $\nu$. However this is the difficulty of all approach of papers \cite{Mitra}-\cite{Alt2015} based upon different asymptotics at $z \to 0$ of $I_{\nu}$ and $I_{-\nu}$ coinciding at $\nu$ integer.

Again in parallel with the simple example of section 2 it is worthwhile to note that nice G-Y formula (\ref{16}) for the ratio of determinants now takes the form:

\begin{equation}
\label{26}
\frac{{\rm Det}_{f_{2}}\hat{D}(p)}{{\rm Det}_{f_{1}}\hat{D}(p)} = \frac{\epsilon \,v'(p\,\epsilon)+ r_{2}v(p\,\epsilon)}{\epsilon \,v'(p\,\epsilon)+ r_{1}v(p\,\epsilon)},
\end{equation}
and it exactly coincides with ratio (\ref{22}) if we use in (\ref{26}) functions determining Green functions (\ref{7}), that is if it is taken $v(pz)=z^{d/2}K_{\nu}(pz)$ and $r_{1(2)}$ are built from $u_{f} = z^{d/2}[I_{-\nu} + \bar{f}I_{\nu}]$ in a way similar to (\ref{20}):

\begin{equation}
\label{27}
r_{\epsilon\,1(2)} = - \frac{\epsilon \, u'_{f_{1(2)}}(p\,\epsilon)}{u_{f_{1(2)}}(p\,\epsilon)} = -\frac{d}{2} - \frac{\epsilon \,I'_{-\nu}(p\,\epsilon) + \bar{f}_{1(2)} \epsilon \,I'_{\nu}(p\,\epsilon)}{I_{-\nu}(p\,\epsilon) + \bar{f}_{1(2)}I_{\nu}(p\,\epsilon)}.
\end{equation}

\subsection{Two-branes RS-model}

Introduction of the IR-brane at finite $z=L < \infty$ does not make the task of calculation of the one-loop quantum energy too much more complicated than in case of one-brane RS-model considered above. Green function $G_{f-r}^{(L)}(p;z,z')$ satisfying asymptotic b.c. (\ref{3}) at $z \to 0$ and certain Robin b.c. $zG' + rG = 0$ at $z=L$ is given by the expression similar to (\ref{7}) where $z^{d/2}K_{\nu}(pz)$ must be changed to function $v_{r}(pz)$ obeying Robin b.c. $zv' + rv = 0$ at $z=L$:

\begin{eqnarray}
\label{28}
v_{r}(pz)= \frac{\pi}{2\sin \pi\nu}\,z^{d/2}\,[I_{-\nu}(pz) - \gamma_{r}(pL)\,I_{\nu}(pz)],  \qquad   \qquad   \nonumber  \\
\\
\gamma_{r}(pL) = \frac{A_{r}[I_{-\nu}(pL)]}{A_{r}[I_{\nu}(pL)]},  \quad A_{r}[\psi(pz)] = \left(\frac{d}{2} + r \right)\,\psi(pz) + z\,\psi'(pz). \nonumber
\end{eqnarray}
Here for any value of Robin parameter $r$: $\gamma_{r}(pL) \to 1$, $v_{r}(pz) \to z^{d/2}K_{\nu}(pz)$ at $L \to \infty$. Finally Green function $G_{f-r}^{(L)}$ is built from solutions of Eq. (\ref{5}) $v_{r}(pz)$ (\ref{28}) and $u_{f}(pz) = z^{d/2}[I_{-\nu}(pz) + \bar{f}\,I_{\nu}(pz)]$ (like in (\ref{7})):

\begin{eqnarray}
\label{29}
G^{(L)}_{f-r}(p;z,z') = - k^{d-1}\frac{u_{f}(z)\,v_{r}(z')\,\theta (z'-z) + (z \leftrightarrow z')}{u'_{f}v_{r} - u_{f}v'_{r}} =   -\frac{\pi\,k^{d-1}(zz')^{d/2}}{2\sin \pi\nu}\, \cdot    \nonumber
\\ \nonumber
\\
\cdot \,\frac{[I_{-\nu}(pz) + \bar{f}\,I_{\nu}(pz)]\,[I_{-\nu}(pz') - \gamma_{r}(pL)\,I_{\nu}(pz')]\,\theta (z'-z) + (z \leftrightarrow z')}{\gamma_{r}(pL) + \bar{f}(p)},  \qquad
\end{eqnarray}
where $\bar{f}(p)$ and $\gamma_{r}(pL)$ are defined in (\ref{7}) and (\ref{28}).

Thus for $L < \infty$ the looked for ratio of one-loop determinants of differential operator (\ref{5}) determined by two values of parameter $f$ in the double-trace asymptotic condition (\ref{3}) is given by slightly modified B-N formula (\ref{22}):

\begin{equation}
\label{30}
\frac{{\rm Det}_{f_{2}-r}\hat{D}(p)}{{\rm Det}_{f_{1}-r}\hat{D}(p)} = \frac{G^{(L)}_{f_{1}-r}(p;\epsilon,\epsilon)}{G^{(L)}_{f_{2}-r}(p;\epsilon,\epsilon)} = \frac{I_{-\nu}(p\,\epsilon) + \bar{f}_{1}(p)I_{\nu}(p\,\epsilon)}{I_{-\nu}(p\,\epsilon) + \bar{f}_{2}(p)I_{\nu}(p\,\epsilon)}\cdot \frac{\gamma_{r}(pL) + \bar{f}_{2}(p)}{\gamma_{r}(pL) + \bar{f}_{1}(p)}.
\end{equation}
Surely this expression for ratio of determinants is also given by the RHS of G-Y formula (\ref{26}) if $v_{r}(p\epsilon)$ from (\ref{28}) and $r_{\epsilon\,1,2}$ from (\ref{27}) are used in (\ref{26}).

The visible feature of expression (\ref{30}) is that its RHS includes two factors: one depending only on $\epsilon$ and the other one depending only on $L$. Therefore one-loop vacuum energy $V^{(d)}_{f_{2}-r} - V^{(d)}_{f_{1}-r}$ corresponding to ratio (\ref{30}) and given by standard expression (\ref{9}) consists of two terms depending on $\epsilon$ and on $L$. In particular taking in (\ref{30}) $f_{2} = \infty$ and $f_{1} = 0$ the following formula for difference of regular and irregular one-loop quantum energies is obtained in two-branes RS-model:

\begin{eqnarray}
\label{31}
V^{(d)}_{+(L)} - V^{(d)}_{-(L)} = \frac{\Omega_{d-1}}{2(2\pi)^{d}}\,\frac{1}{\epsilon^{d}}\,\int_{0}^{\infty}\,y^{d-1}dy\,\ln\left[\frac{I_{-\nu}(y)}{I_{\nu}(y)}\right] - \nonumber  \\
\\
\frac{\Omega_{d-1}}{2(2\pi)^{d}}\,\frac{1}{L^{d}}\,\int_{0}^{\infty}\,y^{d-1}dy\,\ln\left[\frac{\left(\frac{d}{2}+r\right)
I_{-\nu}(y)+y\,I_{-\nu}'(y)}{\left(\frac{d}{2}+r\right)I_{\nu}(y)+y\,I_{\nu}'(y)}\right].     \nonumber
\end{eqnarray}
In receiving (\ref{31}) from general formula (\ref{30}) the definition of $\gamma_{r}(pL)$ given in (\ref{28}) was used.

It is instructive to compare this result with one-loop quantum energy in RS-model calculated in \cite{Goldb}, \cite{Pomarol} where not asymptotic b.c. (\ref{3}) but Robin b.c. with fixed Robin coefficient is imposed at $z=\epsilon$. Then, as it is shown in \cite{Goldb}, \cite{Pomarol}, UV-finite non-local term of the one-loop quantum potential calculated for integer $\nu$ includes dependence on $\ln(L/\epsilon)$, hence it gives hope for dynamical explanation of the large mass hierarchy. Nothing of this kind is present in expression (\ref{31}). That is one-loop potential calculated for asymptotic b.c. can not serve a tool of stabilization of IR-brane.

\section{Conclusion: some tasks for future}

\quad The main message of this paper perhaps may be expressed in one word "simplicity". The surprising simplicity of Barvinsky-Nesterov (B-N) and Gelfand-Yaglom (G-Y) methods of calculation of quantum determinants hopefully opens new possibilities in studying quantum effects in higher dimensional models.

In particular one-loop potential (\ref{24}) as a function of double-trace parameter $f$ may be of the Coleman-Weinberg type in certain Schwinger-Dyson gap equation determining $f$ self-consistently.

However, interesting results in this direction of thought may be expected for integer $\nu$ when formulas of the paper can not be applied directly because $I_{\nu}=I_{-\nu}$ in this case. For $\nu$ integer Green function of differential operator $\hat{D}(p)$ (\ref{5}) may be easily constructed from $z^{d/2}I_{\nu}(pz)$ and $z^{d/2}K_{\nu}(pz)$. Here the problem is in the lack of physically motivated analogy of asymptotic expression (\ref{2}) when $\nu$ is integer. Hence it is not clear what may be the dependence $\bar{f}(p)$ in (\ref{7}) introduced in \cite{Mitra} in case of non-integer $\nu$. Meanwhile function $\bar{f}(p)$ essentially determines the form of the physically important potential (\ref{24}) of the double-trace parameter $f$.

Another possible field of future studying is construction of Schwinger-DeWitt (S-DW) expansion in RS-model on the basis of B-N or G-Y methods applied in this paper. In \cite{Alt2015} it was shown that in one-brane RS-model S-DW expansion for curvature in $d$-space is plagued by IR-divergencies in higher terms of the expansion, and that these divergencies are regularized in two-branes RS-model, that is when $L < \infty$. The same role of the term depending on $L$ in expression (\ref{31}) may be expected. This is the question for future research.

\section*{Acknowledgements} Author is grateful for fruitful discussions to Andrei Barvinsky, Ruslan Metsaev, Dmitry Nesterov, Igor Tyutin, Mikhail Vasiliev, Boris Voronov and other participants of Seminar in the Theoretical Physic Department of P.N. Lebedev Physical Institute.

\section*{Appendix. G-Y formula and Casimir effect in one line}

Short explanation of the Gelfand-Yaglom approach was given in the Introduction. Here we demonstrate on some physical problems that G-Y method immediately gives familiar results obtained conventionally in a rather lengthy way. Examples considered below refer to flat $(d+1)$-dimensional space and to elementary differential operator $\hat{D}_{0}(p)$ (\ref{10}).

1. Classical Dirichlet-Dirichlet problem ($0 < z < L$): $\phi(0)=0$, $\phi(L)=0$. $v=C\cdot \sinh(pz)$ is a solution of Eq. $\hat{D}_{0}\phi = 0$ satisfying b.c. at $z=0$. Then according to G-Y method "Dirichlet-Dirichlet" determinant ${\rm Det}_{D-D}\hat{D}_{0} \sim \sinh(pL)$. This yields expression for quantum potential in $d$ dimensions:

\begin{equation}
\label{32}
V_{D-D}^{(d)} = \frac{1}{2}\int \frac{d^{d}p}{(2\pi)^{d}}\,\ln\,[\sinh(pL)] = A + BL - \frac{1}{L^{d}}\frac{\Omega_{d-1}}{(2\pi)^{d}\,2^{d+1}\,d}\,\int_{0}^{\infty}\frac{y^{d}dy}{e^{y}-1},
\end{equation}
where volume of sphere of unit radius of dimension zero must be taken equal to 2 ($\Omega_{1-1} = 2$); $A$, $B$ are irrelevant divergent constants. Last term in (\ref{32}) which is UV-finite and tends to zero at $L \to \infty$ is Casimir potential $V_{Cas\, D-D}^{(d)}$. It is easy to check that (\ref{32}) gives its well known \cite{Mostep} values in (1+1) and in (3+1) dimensions: $V_{Cas\, D-D}^{(1)}L= - \pi/24$, $V_{Cas \, D-D}^{(3)}L^{3}= - \pi^{2}/1440$ (for electromagnetic field this result must be multiplied by 2 - number of polarizations of e-m field).

2. In the same way Casimir potential may be calculated in Dirichlet-Neumann problem (${\rm Det}_{D-N}\hat{D}_{0}(p) \sim \cosh pL$) and in many other problems. One of the striking examples of power of G-Y method is calculation of Casimir potential in $M^{d}\times S^{1}$ when $z$ is a circle of length $L=2\pi\rho$. In this case spectra of periodic (untwisted) or antiperiodic (twisted) modes are found from equations $\cos (\sqrt{\lambda_{n}- p^{2}}\,L) = \pm 1$. Then  according to G-Y, in untwisted case for example, ${\rm Det}_{\it{untw}}\hat{D}(p) \sim (\cosh pL - 1)$ (${\rm Det}_{\it {tw}}\hat{D}(p)\sim (\cosh pL + 1)$ for twisted modes). UV-finite term of vacuum energy

\begin{equation}
\label{33}
V_{\it{untw}}^{(d)} = \frac{1}{2}\int \frac{d^{d}p}{(2\pi)^{d}}\,\ln\,[\cosh(pL) - 1] = A + BL - \frac{1}{L^{d}}\frac{\Omega_{d-1}}{(2\pi)^{d}\,\,d}\,\int_{0}^{\infty}\frac{y^{d}dy}{e^{y}-1},
\end{equation}
gives in particular well known results for $d=1$: $V_{Cas\,{\it{untw}}}= 4 V_{Cas\,D-D}$ - cf. (\ref{33}) and (\ref{32}), and for Casimir effect on torus in 5 dimensions, i.e. for $d=4$: $V^{(4)}_{Cas\, {\it{untw}}} \cdot \rho^{4} = - 3\zeta(5) / (2\pi)^{6}$, received in \cite{Mostep}, \cite{Candelas} with rather complex calculations. It is easy to get in the same way well known values of Casimir potential for twisted modes.

3. G-Y method also gives at once final formula for Casimir potential in case of general mixed Robin b.c. imposed on solutions of Eq. $\hat{D}_{0}\phi = \lambda \phi$ on both borders $z=a$ and $z=b$ (prime means derivative over $z$):

\begin{equation}
\label{34}
\phi'(a) + r_{a}\phi(a)=0; \,\, \, \phi'(b) + r_{b}\phi(b) = 0,
\end{equation}
where $r_{a,b}$ are Robin "masses". Function $v(z) = r_{a}\sinh p(z-a) - p\cosh p(z-a)$ is a solution of homogeneous equation $\hat{D}_{0}v = 0$ obeying Robin b.c. at $z=a$. Then G-Y method says that ${\rm Det}\hat{D}_{0} \sim (v'(b) + r_{b}v(b))$. This gives straight away for the Casimir potential (which is UV-finite term of $V = \int\,d^{d}p\,\ln[v'(b) + r_{b}v(b)]$) expression identically coinciding with the massless version of formula (22) of paper \cite{Elizade} (after substitutions $p \to x$, $b-a \to a$, $r_{a} \to \beta_{2}^{-1}$, $r_{b} \to \beta_{1}^{-1}$).

The generalization of the above formulas for the case of a massive scalar field is obvious.

\end{document}